# Performance Characterization of Heliotrope Solar Hot-Air Balloons during Multihour Stratospheric Flights

Taylor D. Swaim[a], Emalee Hough[a], Zachary Yap[a], Jamey D. Jacob[a], Siddharth Krishnamoorthy[b], Daniel C. Bowman[c], Léo Martire[b], Attila Komjathy[b], and Brian R. Elbing[a,*]

[a] *Oklahoma State University, Stillwater, OK, USA*
[b] *Jet Propulsion Laboratory, California Institute of Technology, Pasadena, CA, USA*
[c] *Sandia National Laboratories, Albuquerque, NM, USA*

*Corresponding author:* Brian R. Elbing, elbing@okstate.edu

## Abstract

Heliotropes are passive solar hot air balloons that are capable of achieving nearly level flight within the lower stratosphere for several hours. These inexpensive flight platforms enable stratospheric sensing with high-cadence enabled by the low cost to manufacture, but their performance has not yet been assessed systematically. During July to September of 2021, 29 heliotropes were successfully launched from Oklahoma and achieved float altitude as part of the Balloon-based Acoustic Seismology Study (BASS). All of the heliotrope envelopes were nearly identical with only minor variations to the flight line throughout the campaign. Flight data collected during this campaign comprise a large sample to characterize the typical heliotrope flight behavior during launch, ascent, float, and descent. Each flight stage is characterized, dependence on various parameters is quantified, and a discussion of nominal and anomalous flights is provided.

# 1  Introduction

The heliotrope (Bowman et al., 2020) is a passive solar hot-air balloon capable of delivering scientific payloads to the upper-troposphere and lower-stratosphere for several hours at a nearly constant altitude. It consists of a spherical envelope constructed from 0.31 mil high density polyethylene sheeting (commonly sold as "painter's plastic" at hardware stores). The interior of the balloon was darkened using pyrotechnic-grade air float charcoal powder. This darkened material absorbs sunlight, heating the air inside enough to loft gram to kilogram-scale payloads into the lower stratosphere. Due to their relatively low cost and ease of construction, their usage has seen a rapid growth over the past few years (e.g., Brissaud et al. 2021; Bowman and Krishnamoorthy  2021; Schuler et al. 2022; Silber et al. 2023). Most of these studies have involved a relatively small sample size of heliotropes launched at a specific target of interest. Consequently, insights from comparison of flight dynamics between studies have limited value due to differences in materials, construction procedures, and atmospheric conditions between launch sites. However, during the summer of 2021 a total of 29 heliotropes were successfully launched from Oklahoma as part of the Balloon-based Acoustic Seismology Study (BASS). This paper characterizes the flight stages (ascent, float, and descent) to identify nominal heliotrope flight behavior. This will provide guidance when planning projects that utilize heliotropes as well as establish baseline performance to compare flight dynamics from other studies.

A heliotrope used during the 2021 BASS flight campaign is shown shortly after launch in Figure 1, which shows the balloon envelope (i.e., the balloon material), shroud (i.e., reinforced region near the heliotrope opening), and flight line that included a parachute and two (upper and lower) payloads (i.e., packages that held instrumentation). The typical heliotrope flight begins with filling the envelope with air using an air blower such as an industrial floor drum fan. As sunlight warms the balloon, it eventually becomes buoyant and gradually ascends to a float altitude that is dependent on the flight configuration and atmospheric conditions. Once the balloon reaches float, it remains in nearly level flight for the entire day. Shortly before sunset the balloon begins to descend back to Earth, allowing for recovery of the heliotrope and payloads.

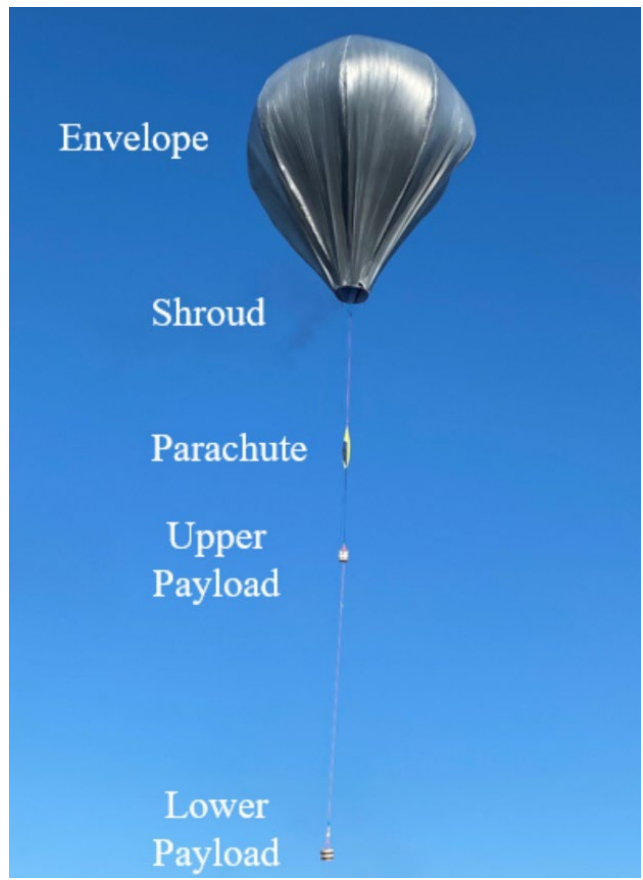

Figure 1. Picture of heliotrope (Flight 16.2) shortly after launch with the envelope, shroud, parachute and payloads (upper and lower) identified.

Development of heliotropes began in 2012, with numerous shapes and sizes tested over the following 3 years. The first well-documented full-day heliotrope flight occurred in 2016, which like the current BASS campaign carried a microbarometer to measure infrasound (i.e., sound at frequencies below human hearing, <20 Hz) (Young et al. 2018). The quasi-spherical shaped balloon envelope was first termed heliotrope during a separate flight campaign in 2017 (Bowman and Albert 2018). Most scientific uses of the heliotrope have been focused on recording acoustic waves and in situ measurements of stratospheric aerosols. The BASS campaign was intended to prove that seismic activity could be detected via acoustic waves propagating into the atmosphere. This has been motivated, in part, to assess the possibility of deploying microbarometers at high altitudes within the Venusian atmosphere as a means of performing seismology on Venus. Direct seismic measurements on Venus have not been made primarily due to surface temperatures of ~460 °C and pressures of nearly 90 atmospheres (e.g., Stofan et al. 1993; Cutts et al. 2015) make sensor performance difficult. However, the Venusian middle atmosphere (45-79 km altitude) is less severe, potentially enabling the use of sensors available today.

Consequently, the primary objective of the 2021 BASS flight campaign was to identify the acoustic signature of natural earthquakes using heliotropes floating in the lower stratosphere. This initiative served as an Earth-analog experiment, simulating the detection of seismic activity on Venus using balloons (Krishnamoorthy and Bowman 2023). However, studying seismicity is not the focus of the current study. Rather, this study leverages these flights to further explore and characterize each stage of the heliotrope flight to better understand the flight platform's performance. Section 2 of the manuscript describes the methods employed, including the envelope and rigging, flight line and instrumentation, flight operations, and overview of flight days. Section 3 characterizes each stage of the flight (launch and ascent, float, and descent and landing). Section 4 discusses the typical nominal flight, observations of anomalous flights. Finally, conclusions are drawn in Section 5.

## 2   Methods

*a. Balloon envelope and rigging*

The basic construction of the heliotropes follows the custom design described in Bowman (2014) and Bowman et al. (2020). However, to meet the additional lift requirements for the current study, the design was scaled up from a 6 m diameter to 7 m. The balloons were manufactured by Big Events Online, Inc. in accordance with the BASS team's specifications. The manufacturer also darkened the inner surface with air float charcoal. However, after the first few launches it was determined that it was best to re-darken the envelope with air-float charcoal the day before launch, which provided darker and more uniform coverage for consistent flight performance. The darkening involved adding charcoal to the envelope and moving it around to coat the entire inner surface. Then any excess charcoal (i.e., charcoal that did not adhere to the envelope) had to be removed before launch to minimize the required lift force. The envelope was formed from 8 gores of either 0.3 mm or 0.7 mm thick plastic with the majority of the flights (all besides the last successful flight day) using the 0.3 mm thick plastic. The thickness increase was done to improve envelope durability and potentially improve the charcoal adherence to the envelope, but there was no significant difference in darkness or performance observed with the thicker plastic. At the top pole the gores were taped together and an additional ~0.5 m plastic disk was applied over the point where the gores meet to reinforce this location. At the bottom pole, a ~0.9 m hole was made for the balloon "mouth" (i.e., balloon opening). Air was added for inflation through the mouth, and the flight line rigging was attached to the balloon at the mouth.

The balloon mouth required reinforcement because it was susceptible to tearing and was the point where the flight line with payloads needed to be attached. For the current project, this was done by taping the shroud ring (consisting of a hula hoop) directly to the mouth opening.

Note that this is not the recommended method as described in Bowman (2014) and Hough et al. (2022), and as a result small tears around the shroud ring did occur during launch. Then from the shroud ring, eight chords of equal length (~1 m) were drawn to the center below the mouth opening and connected with a shackle (e.g., carabiner). The flight line was then attached to the shackle and hung directly below the balloon mouth as shown in Figure 1.

*b. Flight days*

There were a total of 19 days during the 2021 BASS flight campaign where at least one heliotrope launch was attempted. Every attempted launch (successful or failed) is listed in Table 1 with a total of 42 balloon launch attempts with 30 successfully reaching float, though 2 launches had data recording errors and are not included in the analysis. The listed weight ($W$) includes the weight of the payload(s), flight line, and other rigging components. The lighter weight flights ($W = 1.6$ kg) in Table 1 are due to detaching the lower payload during the launch to avoid being caught on an obstacle (e.g., tree, fence, or building). The listed air temperature ($T$), solar irradiance ($Q$), wind speed ($V$), and wind direction (degrees clockwise from due North) are at the time of launch. Ground conditions were measured with the weather station with the exception of the wind direction, which was determined from the initial direction of the heliotrope after launch. The sunrise and sunset times were determined based on the location (i.e., latitude, longitude, and altitude). The sunrise time was calculated based on the location of the heliotrope at launch, while the sunset time required an iterative process due to the westward movement of the floating balloon.

Table 1. Summary of the each flight day with the flight number corresponding to (flight-day).(balloon) and the launch method (G – ground, Ast – helium-assisted, Aug – helium-augmented, F – failed, Err - payload error). Also included is the total flight line weight (*W*), ground conditions at launch (*T* – air temperature, *Q* – solar irradiance, *V* – wind speed, Dir – wind direction with 0° due north), sunrise and sunset time. The sunrise and sunset times are based relative to the balloon location at that time (latitude, longitude, and altitude).

| Date | Flight | Method | W (kg) | T (°C) | Q (W m$^{-2}$) | V (m s$^{-1}$) | Dir (deg) | Sunrise (UTC–5) | Sunset (UTC–5) |
|---|---|---|---|---|---|---|---|---|---|
| 20-Jul | 1.1 | F | | | | | | | |
| 20-Jul | 1.2 | F | | | | | | | |
| 21-Jul | 2.1 | G | 1.6 | NA | NA | NA | NA | 5:53:23 | 21:42:01 |
| 21-Jul | 2.2 | Ast | 3.2 | 19.4 | 133 | 1.3 | -36.9 | 5:53:18 | 21:46:48 |
| 26-Jul | 3.1 | G | 3.2 | 25.6 | 495 | 1.9 | 48.5 | 5:57:29 | 21:29:16 |
| 26-Jul | 3.2 | Ast | 3.2 | 23.0 | 164 | 0.9 | -43.4 | 5:57:45 | 21:29:55 |
| 27-Jul | 4.1 | G | 3.2 | 23.0 | 122 | 0.4 | -28.6 | 5:58:22 | 21:33:24 |
| 27-Jul | 4.2 | G | 3.2 | 25.6 | 246 | 0.4 | -29.9 | 5:58:44 | 21:37:08 |
| 30-Jul | 5.1 | G | 3.2 | NA | NA | NA | NA | 6:01:00 | 21:39:40 |
| 30-Jul | 5.2 | G | 3.2 | NA | NA | NA | NA | 6:01:00 | 21:27:42 |
| 3-Aug | 6.1 | F | | | | | | | |
| 3-Aug | 6.2 | F | | | | | | | |
| 6-Aug | 7.1 | F | | | | | | | |
| 6-Aug | 7.2 | Ast | 3.2 | 24.1 | 418 | 1.7 | 36.7 | 6:08:52 | 21:08:14 |
| 6-Aug | 7.3 | Ast | 3.3 | 25.4 | 489 | 2.0 | 36.7 | 6:08:52 | 21:08:39 |
| 11-Aug | 8.1 | Ast | 3.2 | NA | NA | NA | NA | 6:11:43 | 21:16:35 |
| 11-Aug | 8.2 | Ast | 3.2 | NA | NA | NA | NA | 6:11:43 | 21:08:34 |
| 13-Aug | 9.1 | F | | | | | | | |
| 13-Aug | 9.2 | Ast | 3.2 | 26.3 | 272 | 5.1 | -100 | 6:13:38 | 20:59:56 |
| 17-Aug | 10.1 | F | | | | | | | |
| 17-Aug | 10.2 | Err | | | | | | | |
| 17-Aug | 10.3 | G | 3.2 | NA | NA | NA | NA | 6:17:06 | 21:02:58 |
| 21-Aug | 11.1 | Ast | 3.2 | 23.8 | 184 | 2.3 | -72.8 | 6:20:35 | 20:55:17 |
| 21-Aug | 11.2 | G | 3.2 | 25.2 | 378 | 2.5 | -43.6 | 6:20:33 | 20:52:28 |
| 25-Aug | 12.1 | Err | | | | | | | |
| 25-Aug | 12.2 | Ast | 3.2 | NA | NA | NA | NA | 6:24:02 | 20:53:10 |
| 29-Aug | 13.1 | F | | | | | | | |
| 29-Aug | 13.2 | Ast | 3.2 | 26.2 | 472 | 2.6 | 41.8 | 6:27:27 | 20:28:10 |
| 1-Sep | 14.1 | G | 1.6 | 26.2 | 244 | 1.4 | 30 | 6:29:49 | 20:48:45 |
| 1-Sep | 14.2 | Ast | 3.2 | 27.4 | 354 | 1.2 | 41.6 | 6:29:56 | 20:41:32 |
| 1-Sep | 14.3 | Aug | 1.6 | 27.7 | 414 | 1.1 | 40.5 | 6:29:58 | 20:55:27 |
| 6-Sep | 15.1 | Aug | 3.2 | NA | NA | NA | NA | 6:34:03 | 20:33:14 |
| 6-Sep | 15.2 | F | | | | | | | |
| 6-Sep | 15.3 | Aug | 3.5 | NA | NA | NA | NA | 6:34:05 | 20:26:45 |
| 16-Sep | 16.1 | Aug | 3.3 | NA | NA | NA | NA | 6:24:04 | 20:04:29 |
| 16-Sep | 16.2 | Aug | 3.3 | NA | NA | NA | NA | 6:42:02 | 20:11:44 |
| 22-Sep | 17.1 | Aug | 3.3 | 13.2 | 190 | 1.8 | -121.8 | 6:46:44 | 20:00:20 |
| 22-Sep | 17.2 | F | | | | | | | |
| 23-Sep | 18.1 | Aug | 3.3 | 15.2 | 247 | 1.8 | 17.3 | 6:47:27 | 19:56:20 |
| 23-Sep | 18.2 | Aug | 3.3 | 16.7 | 308 | 2.0 | 13.4 | 6:47:11 | 19:56:25 |
| 27-Sep | 19.1 | F | | | | | | | |
| 27-Sep | 19.2 | F | | | | | | | |

*c. Flight line and instrumentation*

The separation between the balloon mouth and the upper payload was 10 m; a minimum of 3 m is recommended (Bowman et al. 2020) to mitigate flow disturbances from the balloon (e.g., venting of air from the mouth). The final flight configuration (flight days 16-19) used a 1.5 m (5-ft) diameter Rocketman parachute as part of the 10 m separation, as shown in Figure 1. Flight days 1-5 used the naturally deflating balloon envelope to perform the function of a drag skirt. Flight days 6-15 also added a 0.6 m diameter Relationshipware StratoChute nylon parachute above each payload for redundancy until the larger parachutes arrived. The separation between the upper and lower acoustic sensing payloads was 30 m to distinguish the direction of arrival of incident infrasound waves. One flight (Flight 16.2 shown in Figure 1) used a custom-built de-reeler above the lower payload that extended the flight line once above a specified altitude, which reduced the period when the flight line could be caught on obstacles such as trees, power lines, and fences.

Each instrumental payload had a pressure sensor, an inertial measurement unit (IMU), and a satellite tracker packaged within a high density polystyrene foam box. These payloads are similar to those used in previous balloon-based infrasound studies (Krishnamoorthy et al. 2018, 2019, 2020). The pressure sensor was an absolute microbarometer (Digiquartz 6000-15A-IS, Paroscientific) that had a dynamic range of 0 to 103 kPa (15 psia) with a nominal sensitivity of 0.01% of full scale. The IMU was coupled with an external global navigation satellite system (GNSS) in a single Inertial Navigation System (INS) (µINS-IS, InertialSense). The INS units provide a universal time stamp, latitude, longitude, altitude, and three components of velocity sampled at nominally 15 Hz. The Spot TRACE satellite tracker was configured to report its location every 5 minutes while moving as well as a daily report after coming to rest at the end of the flight. This tracker was primarily used for recovery with reports of precise landing coordinates as well as additional movement after landing due to atmospheric, animal, or human causes. In addition, each balloon was also equipped with at least one Automatic Packet Reporting System (APRS) for real-time monitoring of the balloon location and altitude via a web-based interface (https://aprs.fi/). The High Altitude Science StratoTrack APRS radio was secured directly to the flight line approximately 1 m below the upper payload.

*d. Flight operations*

A detailed description of flight operations is provided in Hough et al. (2022), but a brief overview of critical details is provided here for completeness. Flight days were identified based on weather forecasts and projected trajectories. The ideal weather conditions were low winds (< 2.2 m/s) and clear skies. The projected trajectories were used to assess if the flights were likely to pass over the region of interest (i.e., region of significant seismic activity located in central Oklahoma), the distance required to recover the payloads, and the likelihood of landing in a densely populated area. The trajectories were regularly updated based on forecasted wind conditions until launch. On a few occasions, a launch was scrubbed at the last minute because either the balloon trajectories or launch conditions changed such that they were no longer favorable. The day before the launch all equipment was gathered and a Notice to Airmen (NOTAM) for an unmanned free balloon was issued by the Federal Aviation Administration (FAA).

Three different launch sites were used during the 2021 BASS flight campaign; an Oklahoma Mesonet (Brock et al. 1995; McPherson et al. 2007) site (36.064, -97.213), the OSU Unmanned Flight Station (36.162, -96.836), and a cow pasture adjacent to the flight station (36.162, -96.832). The flight station was relatively flat with a well-maintained terrain (i.e., regularly mowed). It had nominally 380 m by 90 m of unobstructed ground that was surrounded by barbed wire fences and buildings. The fences and building posed a challenge when launching in less-than-ideal conditions. The cow pasture and Mesonet site had fewer obstacles with much greater distances between fences and trees. However, tall grass and uneven ground made these launches challenging as well. These sites were selected primarily due to ease of access and sufficient unobstructed space to walk/run the heliotropes while they were gaining lift.

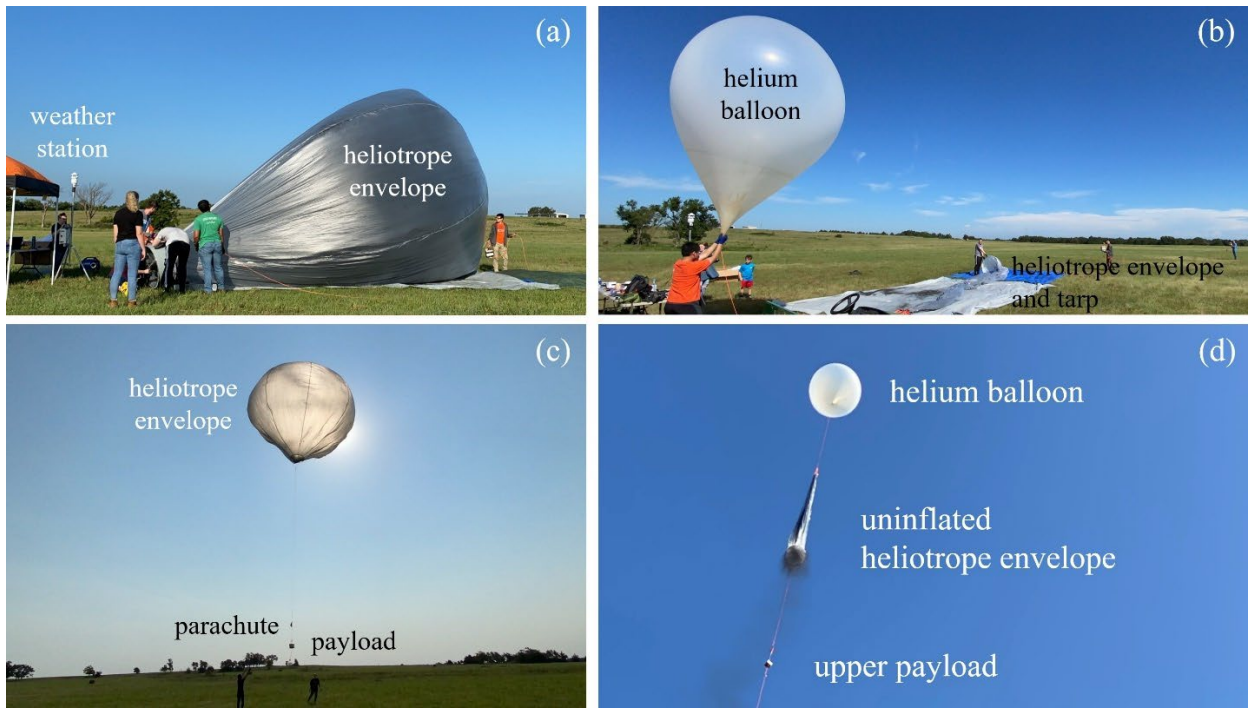

Figure 2. Pictures of (a,c) traditional ground and (b,d) helium-assisted launches (a,b) just prior to launch and (c,d) after the envelope was airborne.

On each day, a specific spot was selected for filling the envelope to maximize the unobstructed field in the direction of the wind while maintaining sufficient space (i.e., minimum of 10 balloon diameters) around the balloon if the wind were to shift. A tarp was spread out at this filling site, which was used to prevent tearing of the envelope while in contact with the ground (see Figure 2a). Near the filling site a telemetry and weather station was set up. Ground conditions (wind speed and direction, solar irradiance, temperature, etc.) were carefully monitored using a Lufft WS600-UMB Weather Sensor weather station. The telemetry station was used to confirm the proper operation of all tracking devices immediately prior to launch as well as confirming operation of the instrumentation. Power was provided at the filling site by extension cords or a generator to power the telemetry station and a fan used to fill the envelope during traditional launches. Helium tanks were always available at the launch site in case an unconventional launch (i.e., any launch other than a ground launch as described in Bowman et al. (2020)) was required. The launch would occur once the solar irradiation was sufficiently high (typically ~190 W/m$^2$).

Three different launch methods were used depending on the ground conditions, which are termed ground launch, helium-assisted, and helium augmentation. Ground and helium-assisted launch methods are pictured in Figure 2, which were both utilized in Bowman et al. (2020) while the third was developed during this flight campaign. A ground launch was the preferred method because it does not require any helium. For a ground launch, the envelope was inflated by blowing air into it with a tiltable industrial floor drum fan. If there was wind, the inflated heliotrope was walked (or ran) as the solar radiation heats the air inside the balloon and gradually became buoyant. This method requires relatively clear skies and low winds. During this campaign, successful ground launches typically required wind speeds below ~1.8 m s$^{-1}$ with a solar irradiance level above 170 W m$^{-2}$. However, variations were observed based on the launch site terrain, length of the unobstructed path in the direction of the wind, and the length of the flight line. The helium-assisted launch method was utilized when launch conditions were unfavorable with high winds and/or significant cloud coverage. In addition, a Saharan dust cloud was over Oklahoma at the start of the campaign forcing helium-assisted launches on some clear mornings and two of the failed launches (Flights 1.1 and 1.2). For this method, a helium balloon was attached to the top of an uninflated envelope, which would rapidly lift the heliotrope away from the ground. Depending on the helium balloon quality and the amount of helium, the helium balloon would burst at an altitude between 12 and 23 km. The heliotrope envelope would then be ram inflated as it descended after the helium balloon burst. The sun would then heat the air within the envelope, and the heliotrope would either ascend or descend to its stable float altitude.

The conclusion of Bowman et al. (2020) noted that a critical need was to develop launch techniques that can tolerate windier conditions. This motivated the development of the third method (helium augmented) utilized during this campaign. This method fills the envelope with air from a fan, but prior to walking the heliotrope, helium was pumped directly into the heliotrope envelope. This gave the heliotrope some instantaneous lift that significantly reduced the period of time when it was inflated and near the ground, where most failures occurred. This method used significantly less helium than the helium-assisted launches and was utilized (once discovered) when winds were slightly higher than ideal for ground launches. More details about the launching of heliotropes are provided in Hough et al. (2022).

Once launched, the heliotrope was tracked in real-time using the APRS tracker as previously described. Since APRS relies on line-of-sight with ground receivers, the signal was often lost in the final few hundred meters of descent. Thus, the SPOT TRACE satellite tracker was used to determine the final landing location. A detailed discussion of recovery procedures is provided in Hough et al. (2022). Every launched payload was successfully recovered during the 2021 BASS campaign.

## 3  Flight Characterization

*a. Overview of flight stages*

Following the convention defined in Bowman et al. (2020), a heliotrope balloon flight is generally divided in to three distinct flight stages; ascent, float or neutral buoyancy, and descent (see Figure 3). In the current work, the ascent stage began once the last payload on the flight line was airborne (i.e., the conclusion of the launch). This time was determined from manual inspection of the altitude data, which typically had a decrease in rise rate (or even a small drop in altitude) when the final payload would lift. For ground and helium augmented launches, the end of the ascent stage was defined as when the vertical velocity of the heliotrope dropped below 1.2 m s$^{-1}$. For a helium-assisted launch, the ascent ended once the helium balloon burst. Several factors determine when the helium balloon would burst with the primary one being the pressure of the filled helium balloon. Secondary factors include UV damage to the helium balloon prior to launch and the balloon quality (e.g., material uniformity). Consequently, there was significant variation in the burst altitude, which resulted in the burst occurring above (type-1) or below (type-2) the final float altitude. The float or neutral buoyancy stage was identified based on the altitude achieving a relatively constant value and the average vertical velocity dropping to close to zero. These definitions create a transitional stage between the ascent and float stages. The end of the float stage was similarly identified based on the deviation from a relatively constant altitude and near zero average vertical velocity. Of note, the IMU power was typically lost prior to descent (see Figure 3), which required this stage to be defined with the lower resolution APRS tracker. However, both the IMU and APRS were used to identify the ascent and float stages, which produced very similar results.

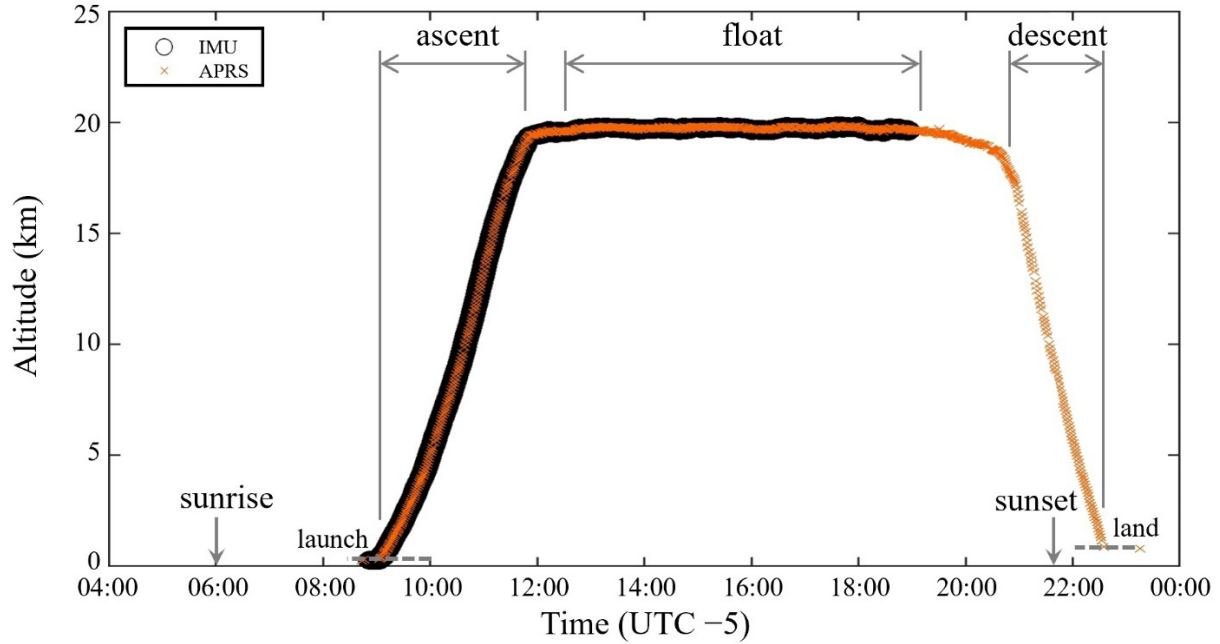

Figure 3. Altitude data from the upper payload IMU and APRS tracker for a typical flight (Flight 4.2) illustrating the flight stages (ascent, float, and descent). Altitudes at launch and landing are marked with horizontal dashed lines. The IMU has a higher sampling rate than the APRS (i.e., the data points are so close it appears as a thick line) but loses power before the end of the float stage. Lines denoting the start and end of stages as well as sunrise/sunset are also included for reference.

The descent stage was defined as beginning once the vertical velocity decreased below -2 m s$^{-1}$. However, the heliotrope would slowly descent prior to the start of the descent stage, as seen in Figure 3. This was due to the solar irradiance level decreasing as sunset approached because of the increasing path length within the atmosphere. The heliotrope would continue to descend until the conclusion of the flight when the payloads landed on the ground. The APRS tracker typically lost signal approximately 200 m above the ground, from which the APRS data was linearly extrapolated to the final landing elevation identified by the satellite tracker to approximate the end of descent (i.e., landing time).

## b. Launch and ascent

The 2021 BASS flight campaign had 10 ground, 8 helium-augmented, 5 helium-assisted (burst above float altitude), and 6 helium-assisted (burst below float altitude) launches that successfully reached float with data. The average speeds at take-off and during ascent as well as the total ascent time for the different launch methods are summarized in Table 2. The average take-off speed ($V_{to}$) was determined from the ascent rate between launch and when the heliotrope was 50 m above the ground. Comparison between ground and helium-augmented launches shows that the added helium increased $V_{to}$ by 0.68 m s$^{-1}$ on average, which is a 126% increase in take-off velocity. This significantly reduces the period of time when the flight line was at the height of local objects (i.e., trees, power lines, etc.). Consequently, there were no failed helium-augmented launches due to the flight line catching on an object. The variability (i.e., standard deviation) in the take-off speed increased with the helium, which is likely related to the lack of a precise method for adding a given volume of helium to the envelope.

Table 2. Average takeoff speed ($V_{to}$), average ascent rate ($V_a$), and time to float ($t_f$) for each launch method. Only flights with two payloads were included.

|  | $V_{to}$ (m s$^{-1}$) | | $V_a$ (m s$^{-1}$) | | $t_f$ (s) | |
| --- | --- | --- | --- | --- | --- | --- |
|  | avg | std | avg | std | avg | std |
| Ground | 0.54 | 0.17 | 1.8 | 0.14 | 13,400 | 1,100 |
| Helium-Augmented | 1.22 | 0.30 | 2.1 | 0.17 | 11,800 | 600 |
| Helium-Assisted (burst above) | 4.2 | 0.89 | 5.4 | 1.3 | 6,200 | 1,700 |
| Helium-Assisted (burst below) | 4.5 | 0.74 | 3.9 | 0.83 | 7,100 | 1,300 |

The helium-assisted launches had different take-off velocities with type-2 (burst below float) being on average 0.3 m s$^{-1}$ faster than type-1 (burst above float). This was primarily due to type-1 (burst above) having less helium resulting in lower net lift force and a higher altitude before bursting. The type-2 (burst below) ascents had more helium causing it to be over pressurized at a lower altitude and bursting prematurely. The average burst altitude of helium-assisted type-1 (burst above) and type-2 (burst below) were 22,500 ± 1,760 m and 14,300 ± 5,780 m, respectively, at 2 standard deviations. The type-2 (burst below) launches rapidly ascend until the helium balloon bursts, the heliotrope ram inflates, and slowly ascends to float altitude. In contrast, type-1 (burst above) ascents traveled at a steady speed to an altitude above float altitude before the helium balloon bursts, which ends the ascent stage for this type. Type-2 (burst below) launches also had a drop in altitude after the burst, but it was not as large as type-1 (see Figure 4).

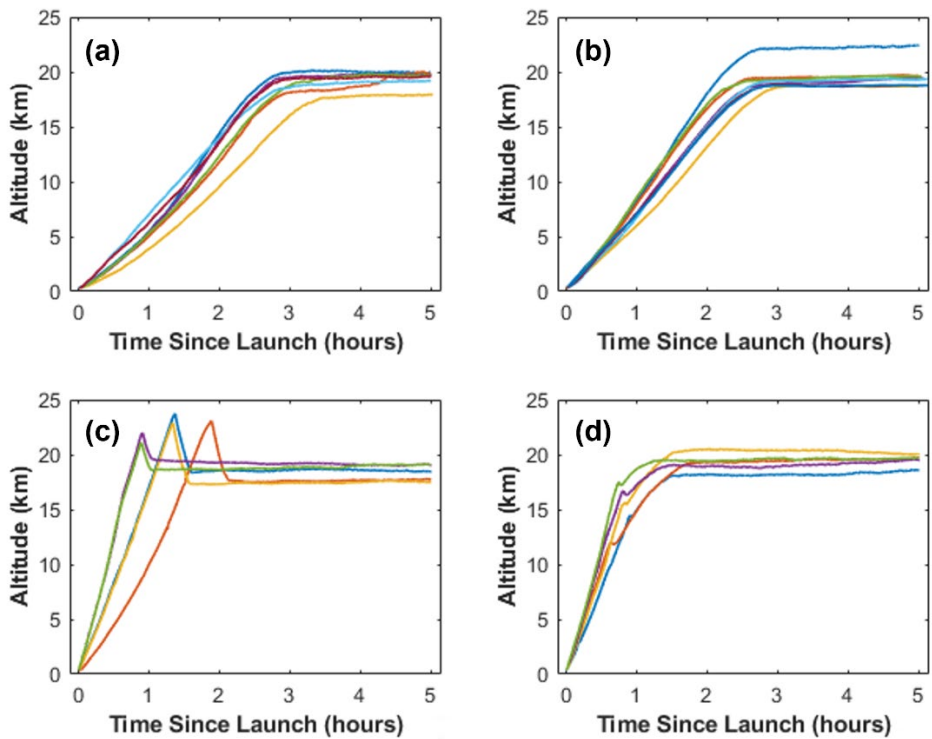

Figure 4. Launch sequence for (a) ground, (b) helium-augmented, (c) type-1 (burst above) helium-assisted, and (d) type-2 (burst below) helium-assisted. Time is measured relative the given flights launch time. Ground launches exhibit the slowest ascent profile, while helium-augmented launches showcase a similar profile with an initial higher ascent rate. The helium-assisted launches are divided into type-1 (helium balloon burst above float altitude) and type-2 (helium balloon burst below float altitude).

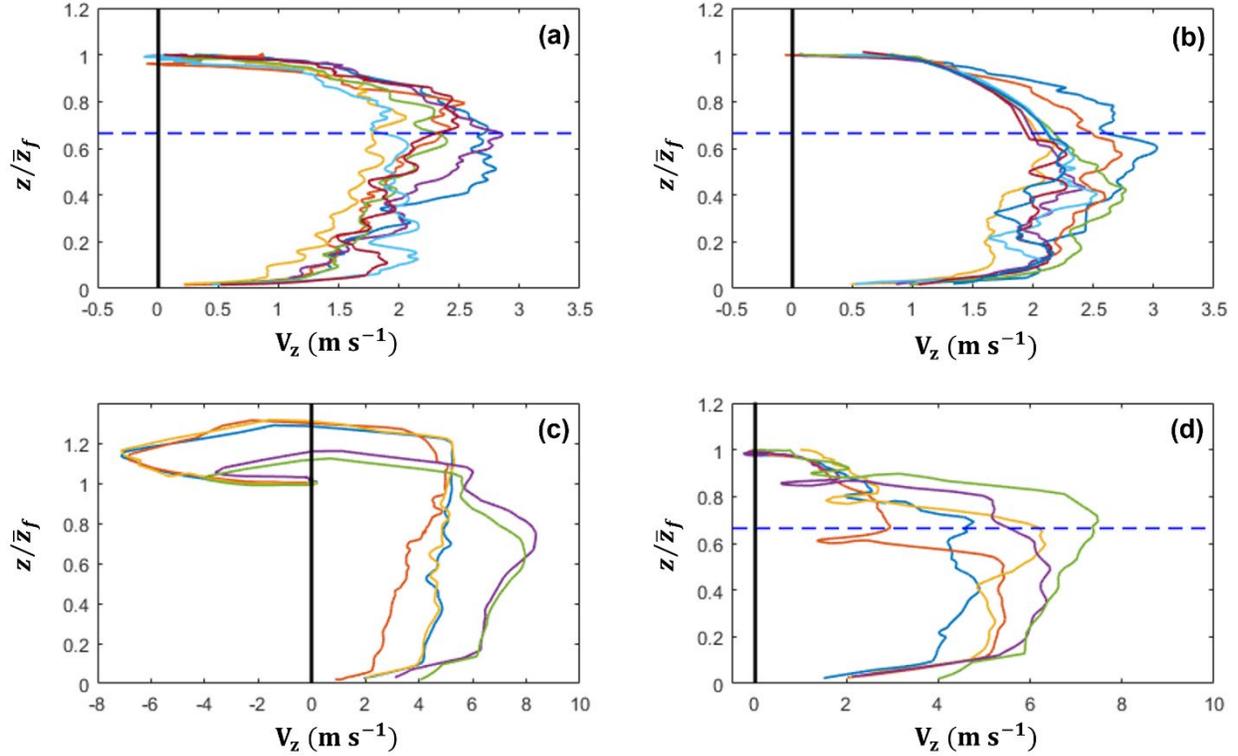

Figure 5. The altitude (*z*) normalized by the average float altitude $(\bar{z}_f)$ plotted versus the vertical velocity $(V_z)$ during ascent for (a) ground, (b) helium augmented, (c) type-1 (burst above float) helium assisted, and (d) type-2 (burst below float) helium assisted launches. The blue dashed horizontal lines mark two-thirds of the float altitude, except for (c) where the helium balloon lifts the balloon above the float altitude.

The vertical velocity $(V_z)$ versus the altitude $(z)$ scaled by the average float altitude $(\bar{z}_f)$ is plotted in Figure 5. Bowman et al. (2020) observed that heliotropes achieve maximum rise rate at approximately two-thirds of their float altitude and that the rise rate fluctuates with an amplitude of up to 1 m s$^{-1}$, which is consistent with the current results independent of the launch method. Variation of the ascent rates was also observed in Bowman et al. (2020), which it was conjectured to be dependent on variations in balloon configuration and/or variability in the darkening of the envelope. The current campaign used nearly the same configuration for each launch with the only significant exceptions being four launches that only had one payload (see Table 1). Some of the ascent rate variability can be attributed to the weight differences. The two ground launches with a payload weight of 1.6 kg (Flights 2.1 and 4.2) had an average ascent rate of 2.1 m s$^{-1}$ ± 0.1 m s$^{-1}$, which was 0.3 m s$^{-1}$ faster than the two payload ground launches that had an average weight of 3.2 kg. Similarly, the one payload helium augmented launch (Flight 14.3) had an average ascent rate of 2.5 m s$^{-1}$, which was 0.4 m s$^{-1}$ faster than the two payload launches (see Table A1). However, variability was still observed in the ascent rate when omitting the flights with a 1.6 kg flight line weight (i.e., single payload launches). There was inconsistency with the level of envelope darkness, especially earlier in the campaign, but no quantitative measurement of the envelope darkness was available.

*c. Float*

The average float altitude of all flights was 19.3 km ± 2.0 km, where the range of variability is twice the standard deviation of the flight averages. The minimum and maximum average float altitudes were 17.4 km and 22.2 km, respectively. While the standard deviation of average was relatively large, the fluctuations during an individual flight were much lower with the standard deviation in altitude during a single flight being 0.2 ± 0.2 km. The single payload (1.6 kg) and double payload (≥ 3.2 kg) flights had an average float altitude of 20.6 ± 2.4 km and 19.1 ± 1.6 km, respectively, with the variation being twice the standard deviation. The flights with one payload (1.6 kg) had average float altitudes that were higher than the two payload (≥ 3.2 kg) flights. However, it should be noted that two of the four single payload flights were on Flight Day 14, which all flights on that day had exceptionally high flights. The three balloons launched on Flight Day 14 achieved the three highest float altitudes of the campaign (20.9 km, 20.9 km, and 22.2 km). Flight 14.2 had two payloads (≥ 3.2 kg) suggesting that this particular day had favorable conditions (e.g., cloud coverage) for higher float altitudes. There also did not appear to be any significant impact on the launch method with the average float altitude for ground, helium-augmented, and helium-assisted launches being 19.5 km, 19.4 km, and 19.0 km, respectively. A summary of the average and standard deviation of the float for each balloon is provided in Table A2.

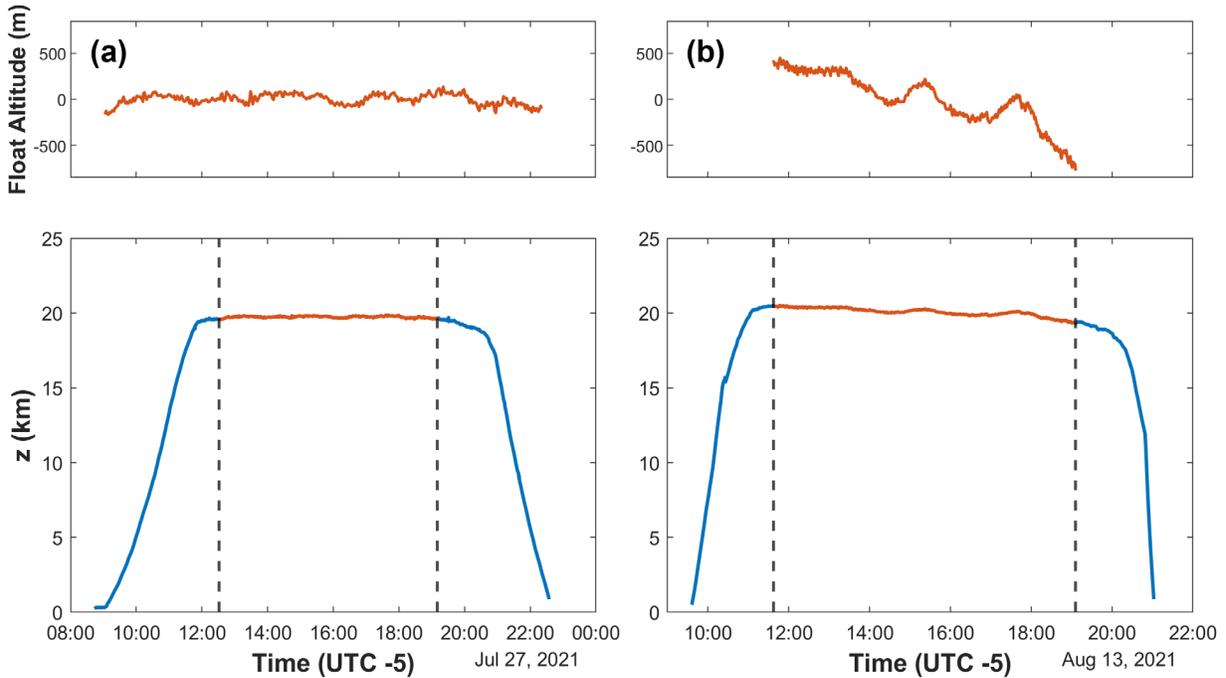

Figure 6. Float behavior examples including (a) Flight 4.2 experiencing a steady float altitude and (b) Flight 9.2 displaying both upward excursions and a gradual decrease in altitude during the float period.

Bowman et al. (2020) identified some of the different behaviors that a heliotrope experiences during the float period; (i) steady altitude, (ii) gradual decrease through the flight, and (iii) upward excursions on the scale of an hour. Bowman et al. (2020) also noted that any combination of these behaviors could be observed even with similar balloons launched at close to the same time. These three behaviors, individually or in combination, were observed in the current study with two examples shown in Figure 6. The gradual decrease behavior can be quantified using the linear slope in altitude during the float period, which the majority of flights had an average slope of approximately -30 m hr$^{-1}$. However, there were a few flights that experienced over -100 m hr$^{-1}$, including Flight 9.2 (included in Figure 6) that had a slope of -129 m hr$^{-1}$. Higher order statistics can also provide a quantitative assessment of these qualitative descriptions. The skewness of the float altitude was negative for nearly every flight, which means that there is a long tail towards lower altitudes. This is also apparent in Figure 6 for the steady altitude flight (Flight 4.2), where most points are at higher altitudes with the fewer lower altitude points having larger deviations from the mean. This is due, in part, to the fact that the altitude consistently had a gradual decline towards the end of the float period as sunset approaches. The average kurtosis, a measure of the tails of the distribution, for all of the flights was 4.3, which a kurtosis of three would be a normal distribution. This indicates that the tails of the distribution are on average slightly heavier than a normal distribution. There were three flights (4.2, 11.1, 11.2) that had significantly larger kurtosis (>8) than the other flights, which this is likely a byproduct of the float stage start and end definition. For example, Flight 4.2 (Figure 6a) appears to still be rising at the start of the float period.

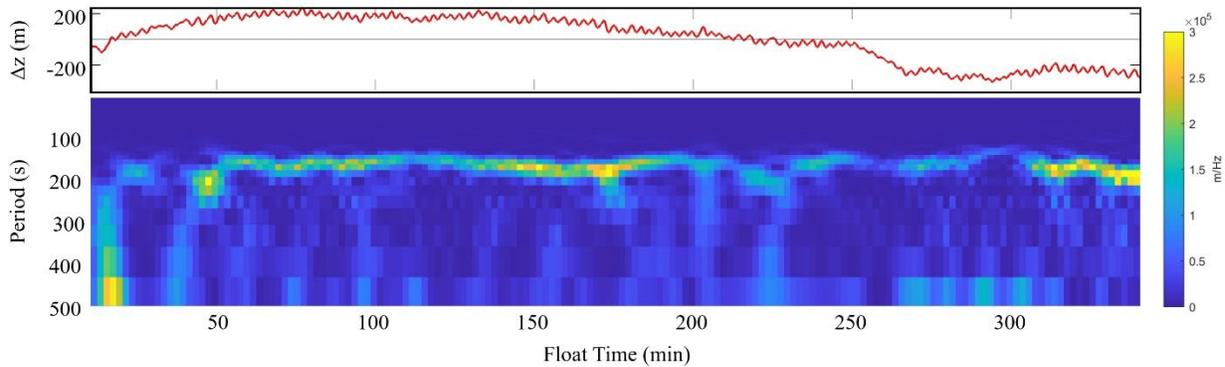

Figure 7. INS data (top) time trace and (bottom) spectrogram of the vertical deviation from the average float altitude ($\Delta z$) from the lower payload during the float period for Flight 16.2. The time is measured relative to the start of the float period. The color indicates the altitude fluctuation amplitude at the corresponding period.

Spectrograms of the altitude during the float stage shows minimal frequency content with periods shorter than ~150 s. Spectral content was observed for every flight with periods longer than ~150 s with most flights showing elevated spectral content between 100 and 200 seconds.

An example, with consistent spectral content within this band and relatively weak spectral content outside of it is shown in Figure 7. Most flights showed larger spectral content with periods longer than 200 s, which at these longer periods the vertical oscillations appeared to be non-linear. Bowman et al. (2020) reported significantly non-linear oscillations with a period centered around 300 s. Bowman et al. (2020) notes that atmospheric gravity waves may induce vertical oscillations in heliotropes, leading to the heliotrope ingesting or exhausting air during the oscillation cycle, contributing to the observed non-linear behavior.

*d. Descent and landing*

As the sun approached the horizon, the heliotrope would begin to descend. Solar irradiance decreased before civil sunset causing the heliotrope to gradually lose lift and consequently altitude. The beginning of the descent stage occurred on average $36 \pm 6$ min before civil sunset at the heliotrope location and altitude. This is shown in Figure 8, which plots the sunset time (based on balloon location and altitude) and the start of the descent stage as a function of the day of the year. The seasonal variation of the sunset during the flight campaign is apparent from the gradual decrease in sunset time. In addition, the start of the descent stage consistently begins ahead of the civil sunset with a weak dependence on the seasonal variation during this period of the year. This was quantified by comparing the slopes of the curve fits for the sunset and start of the descent stage, which the start of descent had a ~6% shallower slope than the sunset.

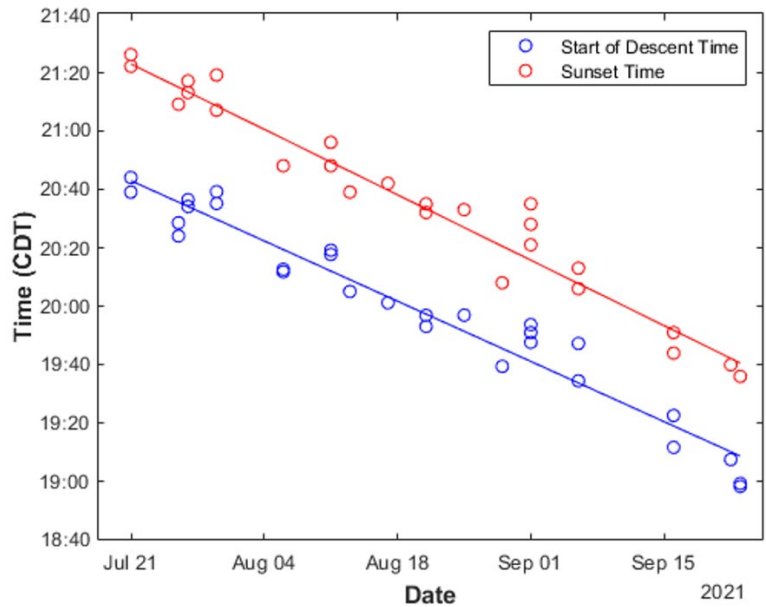

Figure 8. The sunset time (civil twilight at the heliotrope location and altitude) and the start of the descent stage versus the calendar day of the year. The lines are linear best-fit curves to the corresponding data.

During descent, the balloon envelope functioning as a drag skirt was the primary means of slowing the rate of descent and impact speed. However, for a subset of flights (8 total) the envelope detached from the shroud ring during the descent resulting in higher descent rates and impact speeds, which have been classified as "hard landings." Table 3 summarizes the average descent rate, time to landing, and the average impact speed for nominal and hard landings, which the hard landings are grouped based on the additional parachute(s) used. For the 20 nominal descents, the average descent rate was $2.8 \pm 0.3$ m s$^{-1}$ with an average impact speed of $2.2 \pm 0.26$ m s$^{-1}$. For the hard landings, the descent and impact speeds were dependent on the configuration of the redundant parachutes added to the flight line. The vertical variation during the descent for the nominal and hard landing cases are shown in Figure 9.

Table 3. Average descent rate, time to landing, and impact speed for nominal descents and hard landings. There were 20 nominal descents, 1 hard landing with no parachute, 6 hard landings with small parachute(s), and 1 hard landing with a large parachute.

|  | Descent Rate (m s$^{-1}$) | | Time to Landing (min) | | Impact Speed (m s$^{-1}$) | |
| --- | --- | --- | --- | --- | --- | --- |
|  | avg | std | avg | std | avg | std |
| Nominal | 2.8 | 0.3 | 91 | 19 | 2.2 | 0.26 |
| Hard Landing (drag skirt) | 3.7 | – | 30 | – | 27.0 | – |
| Hard Landing (drag skirt and small parachutes) | 3.8 | 0.32 | 29 | 1.3 | 12.0 | 2.2 |
| Hard Landing (drag skirt and large parachute) | 3.2 | – | 41 | – | 6.2 | – |

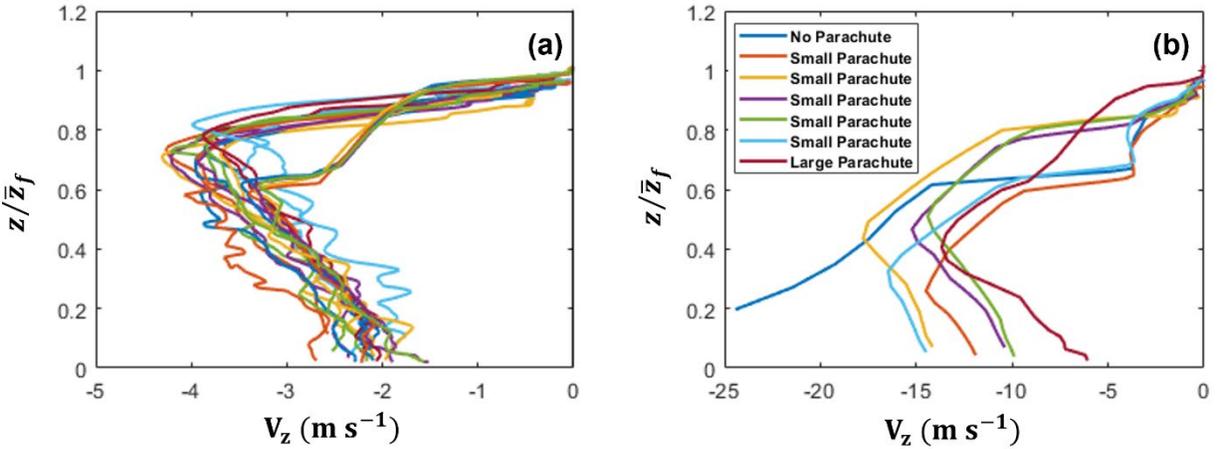

Figure 9. Vertical velocity ($V_z$) during descent plotted versus the altitude ($z$) normalized by the average float altitude ($\bar{z}_f$) for each flight for (a) nominal and (b) hard landings.

# 4 Discussion

*a. Nominal flight behavior*

Nominal flight behavior is determined based on the 28 successful heliotrope flights (9 ground launch, 8 helium-augmented, and 11 helium-assisted) that reached float with data successfully recorded. Ground conditions during the launches ranged in temperature from 13 to 28 °C with an average solar irradiance of ~300 W m$^{-2}$. The average (maximum) wind speed for ground, helium-augmented and helium-assisted launches were 1.3 m s$^{-1}$ (2.5 m s$^{-1}$), 1.7 m s$^{-1}$ (2.0 m s$^{-1}$), and 2.1 m s$^{-1}$ (5.1 m s$^{-1}$), respectively. The average vertical speed at takeoff was nominally 0.5, 1.2, and 4.4 m s$^{-1}$ for ground, helium-augmented, and helium-assisted launches, respectively. The ascent rate increased until reaching a maximum at approximately two-thirds of the float altitude. Helium-augmented launches showed slightly higher ascent rates relative to ground launches prior to the maximum rise rate, but negligible difference at higher altitudes. Helium-assisted ascents are sensitive to the condition and fill of the helium balloon. The average time from launch to float ranged from 104 min (helium-assisted type-1, burst above float) to 223 min (ground) depending on the launch method.

The average float altitude for individual heliotropes ranged from 17.4 km to 22.2 km with an average standard deviation in altitude during the float stage of 0.2 km. There was no apparent dependence between the launch method and the average float altitude, which the average of all flights was 19.3 km. This indicates that the helium-assisted launches did not cause any systematic damage to the envelope and for helium-augmented launches the helium was vented from the envelope prior to reaching float. Higher-order statistics of the float altitude indicates that the tails of the distribution are slightly heavier than a normal distribution and skewed towards lower altitudes. Spectrograms of the altitude during the float stage showed that all flights produced significant spectral content with a period of approximately 150 seconds. Most flights also showed significant, non-linear spectral content for periods longer than 150 seconds.

Descent began on average 36 minutes before civil sunset at the heliotrope location (latitude, longitude, and altitude). The nominal flight had the heliotrope envelope remain attached to the shroud ring and flight line. The average descent rate from these 20 nominal descents was 2.8 m s$^{-1}$ with an average impact speed of 2.2 m s$^{-1}$. It took on average 91 ± 19 min from the beginning of descent to landing, which sets the average time of landing on the ground at nominally 55 minutes after sunset. For non-nominal landings (i.e., hard landings), the impact speed should be controlled with the use of parachutes placed on the flight line.

*b. Anomalous flight observations*

1) RELAUNCHES

Typically, heliotropes are damaged when they land due to the envelope being dragged along as it makes contact prior to stopping. As a result, a landed heliotrope is typically no longer capable of flying without significant repairs. However, during the 2021 BASS flight campaign two heliotropes landed and then relaunched themselves. Flight 3.1 was a traditional ground launch that initially launched at approximately 8:35 AM (UTC-5) and ascended to approximately 458 m above the ground with a relatively constant ascent rate. Then at 8:50 (UTC-5) the ascent ceased and it gradually descended back to the ground. Eye witnesses saw it lift back into the air at 10:41 (UTC-5) with no obvious signs of damage to the envelope. The remainder of Flight 3.1 was nominal. While the standard deviation in the float altitude was relatively high (437 m), the other balloon launched on that day (Flight 3.2) had a larger float altitude standard deviation (452 m).

The second relaunch (Flight 15.3) was an augmented launch with a nominal ascent, float, descent, and landing. The following morning our team contacted the property owner, and they confirmed that the heliotrope had landed with the envelope and both payloads intact, as sketched in Figure 10. However, later that day at 2:59 PM (UTC-5) on 7 Sept 2021 the upper payload satellite tracker reported movement while the lower payload satellite tracker did not. Our team contacted the property owner again, and they were able to confirm that at approximately 3:00 PM (UTC-5) the heliotrope lifted into the air, the lower payload got detached, and then the balloon drifted out of sight. The APRS tracker never reported on the second flight and the payload batteries were exhausted, so there is little information about the second flight. Using the on-board SPOT satellite tracker on the second day shows that the balloon initially moved northward when it relaunched, travelled at least 5.7 km to the northeast, and then turned south. The envelope and upper payload floated south for 113 km, eventually landing in the Caddo National Grasslands of Texas.

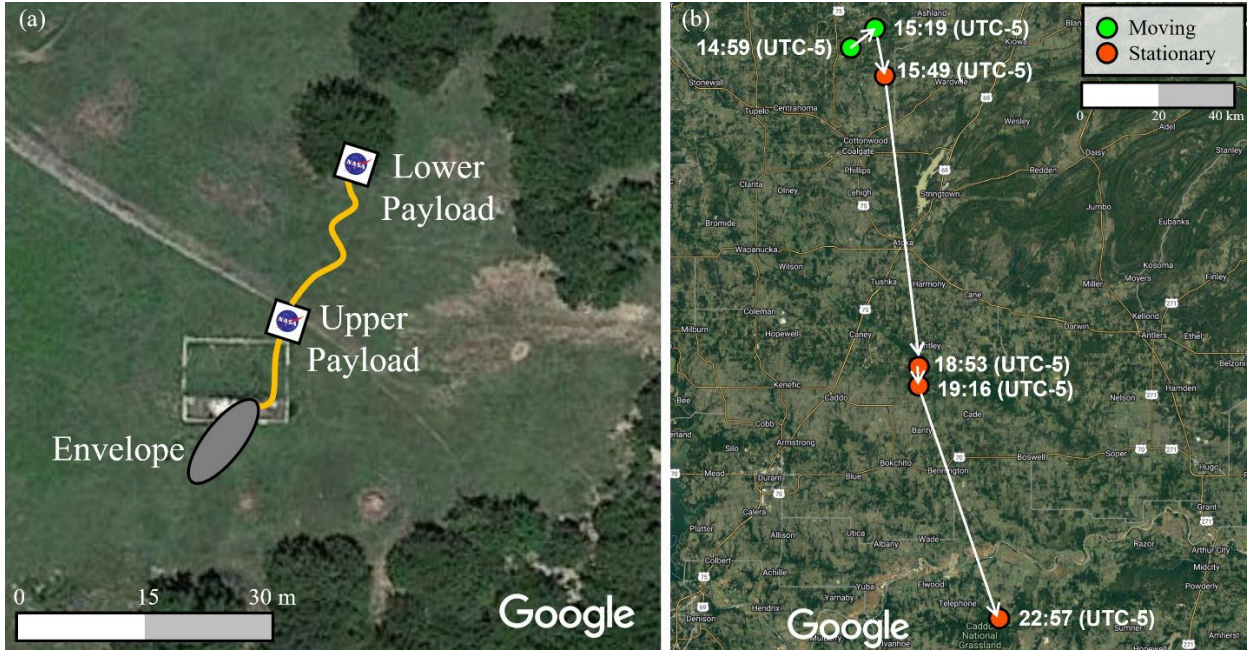

Figure 10. (a) Sketch of the initial landing for Flight 15.3 (envelope and flight line) based on the eyewitness description overlaid on a satellite image (Google). (b) SPOT satellite tracker reports during the second day from the upper payload indicating whether the payload was moving or stationary (Google).

2) UPWARD EXCURSIONS DURING FLOAT

A previous study noted that altitude excursions of 1 km or more occurred when the solar balloon passed over thunderstorms. It was conjectured that the sudden increase in albedo results in an increase in heating and consequently an increase in altitude (Wheeler et al. 2022). During the 2021 BASS flight campaign, Flight 3.2 had a large (≥ 1 km) rise in altitude from its initial relatively level float altitude that lasted over 5 hours. This motivated an investigation of local weather conditions during this flight, which the altitude during the float stage is plotted in Figure 11 along with snapshots of the reflectively from weather radar. This shows that the heliotrope altitude began to increase shortly before storms formed, which would correspond to the period of convective initiation. At 14:50 (UTC-5) the heliotrope was above two storm systems that quickly merged and appears to push the heliotrope around the backside of these eastward moving storms. At 15:40 (UTC-5) NOAA reported significant thunderstorm wind produced from a remnant outflow and a mesoscale convective vortex that helped initiate convection within the area of this storm. Once the heliotrope was west of the eastward moving storm the float altitude returned to nearly that observed during the first 100 min of float. This strongly supports the observations that large excursions ($\geq$ 1 km) being associated with thunderstorms. Of note, the smaller upward excursion shown in Figure 6b was also investigated, and it was over a smaller storm with several storms initiated in the heliotrope region during the float stage.

### 3) HARD LANDINGS

The underlying cause for the 8 hard landings was the detachment of the heliotrope envelope from the flight line. Based on the recovered flight lines, the failure mechanism was the tearing of the envelope plastic from the shroud ring at the heliotrope mouth. Given the uncertainty associated with damage to the recovered envelope and flight line, it is difficult to provide detailed analysis of this failure. However, the altitude of the failure can be nominally estimated from the change in the descent rate as recorded from the APRS tracker. An interesting observation is that all the failures appear to occur before or close to the tropopause, which was nominally estimated from temperature measurements reported from the APRS tracker. A maximum in turbulence occurs near the tropopause (e.g., Söder et al. 2021), which suggests that if the envelope remains intact through the tropopause then it will likely have a nominal landing.

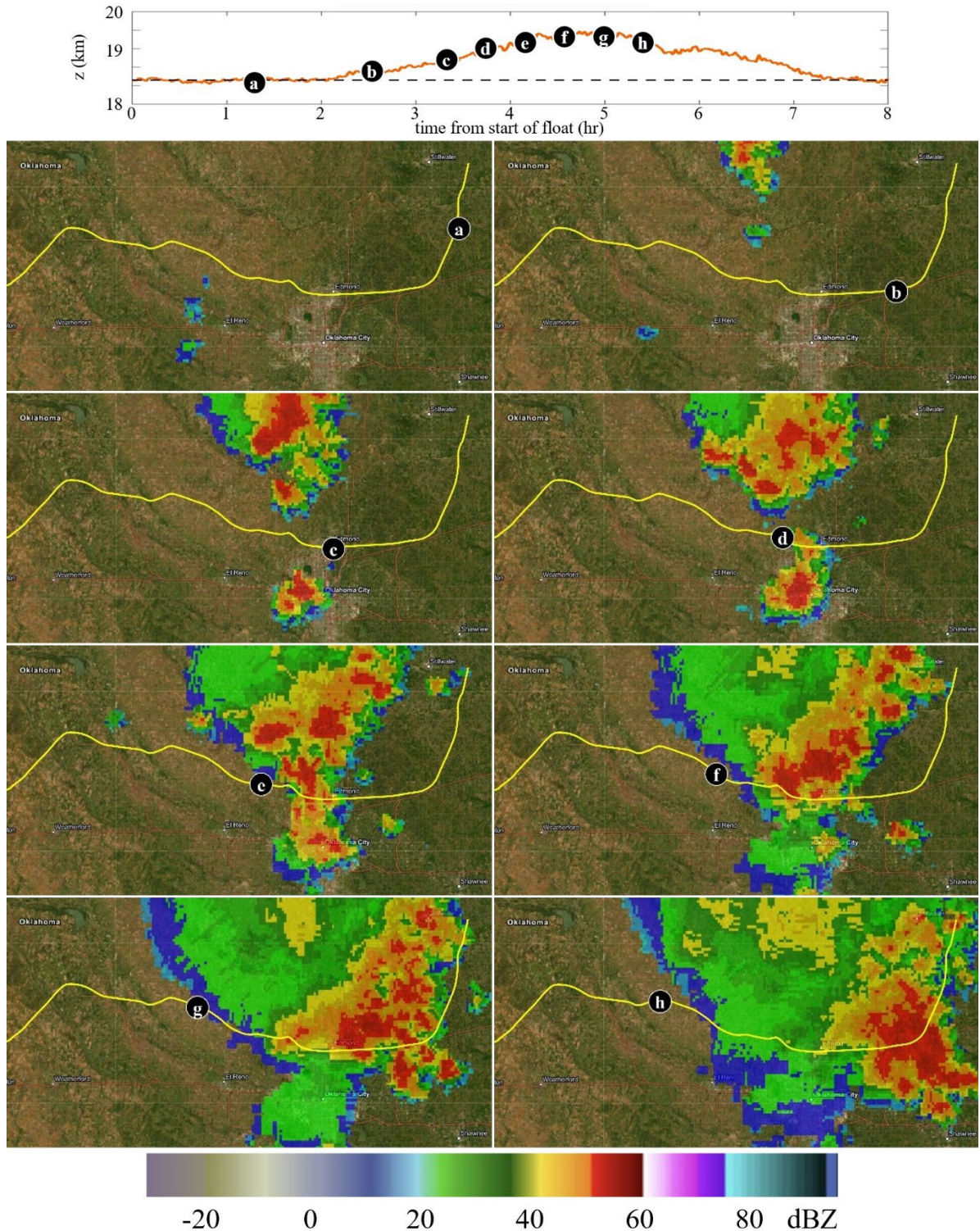

Figure 11. (top) Time trace of Flight 3.2 altitude during the float stage with marked times corresponding to the weather radar images. The dashed line is the average float altitude from the first 100 min of float. (bottom) Weather radar reflectivity at (a) 12:45, (b) 14:00, (c) 14:50, (d) 15:15, (e) 15:40, (f) 16:05, (g) 16:30, and (h) 16:55 (UTC-5) during Flight 3.2 with its trajectory traced in yellow and its current location labeled. Assimilated radar data from NOAA NCEI GIS version 3.1 (www.ncei.noaa.gov/maps/radar).

# 5  Conclusions

The 2021 BASS flight campaign attempted 42 heliotrope launches spread over 19 days with 28 successfully flights. The envelope and flight lines were nearly identical for every flight with the only significant difference being a subset of flights that only had 1 payload instead of two and the use of different launch methods. The heliotropes were launched using either the traditional ground, helium-assisted, or helium augmented methods, which helium-augmented was developed during this flight campaign. The time required to reach float varied significantly depending on the launch method. Helium-augmented launches had a large increase in takeoff speed and a slight increase in the average ascent rate. Besides the takeoff speed and initial ascent rates, there was no significant dependence on the launch method observed through the remainder of the flight. The 7-m diameter heliotropes consistently lifted 3.2 kg payloads to an average altitude of 19.3 km. The float stage showed relatively small fluctuations in altitude during most flights, but upward excursions were observed. Investigation of the largest upward excursion showed that the heliotrope passed over a developing storm system during the period of maximum altitude. The majority of heliotropes gradually descended back to Earth at an average descent rate of -2.8 m s$^{-1}$ beginning on average 36 minutes before civil sunset. All payloads were recovered during the 2021 BASS flight campaign.

The BASS 2021 flight campaign has not only contributed crucial data on heliotrope launches and atmospheric behavior but has also laid a foundation for future experimental planning. This groundbreaking study, the first of its kind, established a robust framework for exploring the effects of different launch methods on heliotrope performance, particularly showcasing the novel helium-augmented approach. The consistent success in reaching the float stage, coupled with the recovery of all payloads, underscores the reliability and resilience of the implemented methodologies. The identification of when the heliotrope was flying over storm systems based on the deviation from the nominal float stage behavior also highlights the potential applications towards atmospheric research. Moving forward, the lessons learned from BASS 2021 will inform and guide the design and execution of future experiments, fostering advancements in our understanding of atmospheric phenomena and enhancing the efficiency and success of subsequent scientific endeavors in this domain.

**Acknowledgements**

The authors would like to thank collaborators at NASA JPL (James Cutts and Michael Pauken) and Sandia National Laboratories (Fransiska Dannemann Dugick and Nora Wynn) who provided onsite assistance with the initial launches as well as guidance throughout the project. We also want to recognize the significant contributions of Alexis Vance, who led the launches done in preparation for this 2021 BASS flight campaign as well as regularly assisted with launches. In addition, we want to thank the large team of students and staff at Oklahoma State University that have helped with preparation, launching, and recovery of heliotropes. This group includes Hannah Linzy, Molly Lammes, Leo Fagge, Reid Williams, Jack Elbing, Trey VanVelsor, Anne Marie Van Dyke, Payton Simmons, Madison Morton, Zachary Morrison, and Landon Dowers, among many others. This research was funded, in part, by the NASA Planetary Science and Technology Analog Research program (NNH19ZDA001N-PSTAR). The storm overflight analysis was supported by the Gordon and Betty Moore Foundation, grant DOI 10.37807/gbmf11559. Contributions from JPL authors were carried out at the Jet Propulsion Laboratory, California Institute of Technology, under a contract with the National Aeronautics and Space Administration (80NM0018D0004).

**Data availability statement**

Data presented in this study are available on the JPL Dataverse (dataverse.jpl.nasa.gov) at the public DOI: 10.48577/jpl.C933HF (Krishnamoorthy, 2024).

# APPENDIX

## Tables of Individual Flight Details

Table A1. Launch and ascent data for all successful flights. The launch/start of ascent was determined from the last IMU sensor on the flight line. *Data averaged between two IMU sensors on same flight.

| Flight | Method | Launch Time (UTC–5) | Launch Altitude (m) | $V_{to}$ (m s$^{-1}$) | End of Ascent Time (UTC–5) | End of Ascent Altitude (m) | Slam Time (UTC–5) | Slam Altitude (m) | Ascent Total Time (min) | Ascent Rate (m s$^{-1}$) |
|---|---|---|---|---|---|---|---|---|---|---|
| 2.1 | G | 8:34:45 | 289 | 0.40 | 11:17:00 | 19300 | | | 162 | 2.14 |
| 2.2 | Ast | 9:43:26 | 288 | 3.15* | 11:05:25* | 24260* | 11:20:35* | 18280* | 82* | 4.98* |
| 3.1 | G | 10:42:11 | 253 | 0.35* | 13:28:00* | 17500* | | | 167* | 1.81* |
| 3.2 | Ast | 9:40:55 | 282 | 3.61* | 11:04:40* | 17640* | 10:36:21* | 14270* | 84* | 3.73* |
| 4.1 | G | 8:22:19 | 281 | 0.29 | 11:30:23 | 16620 | | | 188 | 1.54 |
| 4.2 | G | 9:02:41 | 279 | 0.61 | 11:46:09 | 18990 | | | 163 | 2.07 |
| 5.1 | G | 8:06:30 | 315 | 0.49* | 10:58:53* | 18200* | | | 172* | 1.86* |
| 5.2 | G | 8:44:02 | 282 | 0.69* | 11:35:01* | 18340* | | | 172* | 1.87* |
| 7.2 | Ast | 9:20:10 | 305 | 3.97 | 11:13:35 | 23080 | 11:28:04 | 17650 | 113 | 3.49 |
| 7.3 | Ast | 9:58:17 | 305 | 3.60* | 11:18:22* | 23380* | 11:33:35* | 17280* | 80* | 4.89* |
| 8.1 | Ast | 7:58:16 | 287 | 4.74* | 9:33:09* | 18700* | 8:38:36* | 11750* | 94* | 3.19* |
| 8.2 | Ast | 8:33:19 | 290 | 4.30* | 10:18:19* | 18440* | 9:17:34* | 10580* | 105* | 2.82* |
| 9.2 | Ast | 9:35:15 | 288 | 3.80* | 11:01:59* | 19820* | 10:25:51* | 15360* | 87* | 3.99* |
| 10.3 | G | 8:46:57 | 280 | 0.63* | 11:29:47* | 17910* | | | 162* | 1.91* |
| 11.1 | Ast | 8:21:07 | 285 | 4.92* | 9:34:50* | 18320* | 9:12:02* | 16240* | 74* | 4.37* |
| 11.2 | G | 9:11:32 | 281 | 0.76 | 11:54:16 | 18680 | | | 163 | 1.96 |
| 12.2 | Ast | 10:31:07 | 280 | 4.73* | 11:27:06* | 23980* | 11:30:01* | 17700* | 56* | 6.90* |
| 13.2 | Ast | 9:20:15 | 280 | 5.67 | 10:13:27 | 21090 | 10:22:28 | 18510 | 54 | 6.96 |
| 14.1 | G | 8:37:36 | 276 | 0.99 | – | – | | | – | – |
| 14.2 | Ast | 9:09:40 | 282 | 5.81* | 10:13:49* | 19010* | 9:56:14* | 17100* | 64* | 5.37* |
| 14.3 | Aug | 9:29:22 | 282 | 1.64* | 12:00:12* | 21375* | | | 151* | 2.50* |
| 15.1 | Aug | 8:43:44 | 288 | 1.82* | 11:07:15* | 18790* | | | 144* | 2.32* |
| 15.3 | Aug | 9:17:08 | 289 | 1.01 | 12:01:39* | 17860 | | | 165 | 1.87 |
| 16.1 | Aug | 8:47:14 | 281 | 1.01* | 11:13:44* | 18050* | | | 147* | 2.13* |
| 16.2 | Aug | 9:10:00 | 288 | 1.18* | 11:24:51* | 18490* | | | 135* | 2.40* |
| 17.1 | Aug | 8:20:18 | 290 | 0.87* | 10:50:00* | 18220* | | | 150* | 2.13* |
| 18.1 | Aug | 8:41:22 | 288 | 1.23* | 11:10:09* | 17840* | | | 149* | 2.07* |
| 18.2 | Aug | 9:01:48 | 288 | 1.44* | 11:34:17* | 17960* | | | 153* | 2.02* |

Table A2. Float data for all successful flights. Included is the altitude and time at the start and end of the float stage, the average and standard deviation of the float altitude, and the average and standard deviation in APRS sensor temperature. Only APRS data were used to capture entire float stage. It should be noted, the temperature taken from the APRS was wrapped in a dark colored foam resulting in these temperatures being higher relative to local atmospheric conditions. *APRS sample rate decreased from every minute to every 10 minutes.

| Flight | Method | Start of Float | | End of Float | | Float Altitude | | Sensor Temp | |
|---|---|---|---|---|---|---|---|---|---|
| | | Time (UTC−5) | Altitude (m) | Time (UTC−5) | Altitude (m) | Avg (m) | Std (m) | Avg (°C) | Std (°C) |
| 2.1 | G | 12:06:00 | 20000 | 19:12:00 | 19700 | 19900 | 137 | 3 | 2 |
| 2.2 | Ast | 11:31:00 | 18500 | 19:59:00 | 18000 | 18400 | 166 | -24 | 4 |
| 3.1 | G | 14:16:00 | 18400 | 18:39:00 | 18900 | 19300 | 437 | 14 | 7 |
| 3.2 | Ast | 11:29:30 | 18100 | 19:21:30 | 18100 | 18600 | 452 | 0 | 9 |
| 4.1 | G | 12:08:04 | 17800 | 19:46:04 | 17400 | 17800 | 156 | -2 | 2 |
| 4.2 | G | 12:30:26 | 19600 | 19:09:26 | 19600 | 19700 | 56.9 | 0 | 3 |
| 5.1 | G | 12:19:10 | 19900 | 19:05:09 | 19100 | 19700 | 237 | -4 | 4 |
| 5.2 | G | 12:43:08 | 19700 | 20:16:08 | 18700 | 19400 | 306 | -3 | 3 |
| 7.2 | Ast | 11:57:36 | 17500 | 18:53:36 | 17300 | 17600 | 117 | -5 | 2 |
| 7.3 | Ast | 11:39:46 | 17300 | 19:07:46 | 17100 | 17400 | 140 | -5 | 2 |
| 8.1 | Ast | 10:10:09 | 19300 | 19:03:09 | 19300 | 19700 | 205 | 6 | 5 |
| 8.2 | Ast | 11:09:42 | 19400 | 19:08:42 | 19300 | 19800 | 237 | 7 | 6 |
| 9.2 | Ast | 11:36:01 | 20500 | 19:05:01 | 19300 | 20100 | 278 | 11 | 5 |
| 10.3 | G | 12:14:10 | 18900 | 19:12:10 | 18800 | 19300 | 295 | -11 | 3 |
| 11.1 | Ast | 10:05:54 | 18900 | 18:11:53 | 19600 | 19300 | 250 | 11 | 5 |
| 11.2 | G | 12:28:05 | 19400 | 18:09:05 | 19600 | 19500 | 117 | 13 | 1 |
| 12.2 | Ast | 11:55:57 | 19400 | 19:11:25 | 18400 | 18900 | 283 | -4 | 5 |
| 13.2 | Ast | 10:31:27 | 18600 | 18:50:27 | 18300 | 18800 | 218 | 3 | 5 |
| 14.1 | G | 11:57:39 | 21100 | 19:17:39 | 20600 | 20900 | 142 | 11 | 4 |
| 14.2 | Ast | 10:35:59 | 21000 | 19:07:59 | 20600 | 20900 | 437 | -1 | 2 |
| 14.3 | Aug | 12:26:43 | 22200 | 19:10:43 | 22000 | 22200 | 100 | 1 | 3 |
| 15.1* | Aug | 11:38:30 | 19500 | 18:31:30 | 19300 | 19500 | 94.8 | 13 | 4 |
| 15.3* | Aug | 12:37:13 | 18700 | 18:27:13 | 18400 | 18600 | 112 | 3 | 4 |
| 16.1* | Aug | 12:01:46 | 19000 | 18:41:46 | 18600 | 19100 | 233 | -5 | 4 |
| 16.2* | Aug | 12:22:40 | 19300 | 18:12:40 | 19100 | 19400 | 189 | 5 | 2 |
| 17.1* | Aug | 11:37:37 | 19200 | 18:17:37 | 19000 | 19200 | 162 | 3 | 3 |
| 18.1* | Aug | 12:08:26 | 18700 | 17:48:26 | 18600 | 18700 | 98.0 | 1 | 2 |
| 18.2* | Aug | 12:29:28 | 18700 | 17:49:28 | 18600 | 18700 | 97.3 | 4 | 2 |

Table A3. Descent data for all successful flights with each landing identified as either nominal (Nom) or hard (H). Included is the altitude and time at the start and end of the descent stage, total descent time, average descent rate, and impact speed.

| Flight | Method | Landing Type | Start of Descent Time (UTC–5) | Start of Descent Altitude (m) | Landing Time (UTC–5) | Landing Altitude (m) | Descent Total Time (min) | Descent Avg Rate (m s$^{-1}$) | Impact Speed (m s$^{-1}$) |
|---|---|---|---|---|---|---|---|---|---|
| 2.1 | G | Nom | 20:59:00 | 17296 | 22:36:41 | 920 | 98 | 2.83 | 2.09 |
| 2.2 | Ast | Nom | 21:04:00 | 15814 | 22:20:22 | 1180 | 76 | 3.37 | 2.68 |
| 3.1 | G | Nom | 20:44:00 | 15434 | 22:09:16 | 452 | 85 | 2.89 | 2.11 |
| 3.2 | Ast | Nom | 20:48:30 | 15475 | 22:09:19 | 500 | 81 | 3.08 | 2.44 |
| 4.1 | G | Nom | 20:54:04 | 15318 | 22:12:38 | 814 | 79 | 3.11 | 2.42 |
| 4.2 | G | Nom | 20:56:26 | 17162 | 22:35:06 | 820 | 99 | 2.78 | 2.18 |
| 5.1 | G | Nom | 20:59:09 | 17154 | 22:35:23 | 1000 | 96 | 2.76 | 2.20 |
| 5.2 | G | H | 20:55:08 | 16833 | 21:24:58 | 779 | 30 | 5.62 | 26.9 |
| 7.2 | Ast | Nom | 20:32:36 | 14733 | 21:50:52 | 293 | 78 | 3.11 | 2.42 |
| 7.3 | Ast | Nom | 20:31:46 | 14778 | 21:45:39 | 294 | 74 | 3.34 | 2.73 |
| 8.1 | Ast | Nom | 20:39:09 | 16415 | 22:12:55 | 553 | 94 | 2.85 | 2.14 |
| 8.2 | Ast | H | 20:37:42 | 16611 | 21:11:42 | 509 | 33 | 7.10 | 10.1 |
| 9.2 | Ast | H | 20:25:01 | 16878 | 21:02:39 | 310 | 37 | 6.29 | 12.3 |
| 10.3 | G | Nom | 20:21:10 | 16342 | 21:58:37 | 323 | 97 | 2.76 | 1.98 |
| 11.1 | Ast | H | 20:16:53 | 16145 | 20:34:31 | 604 | 17 | 10.70 | 14.6 |
| 11.2 | G | H | 20:13:05 | 16952 | 20:40:11 | 512 | 26 | 8.37 | 10.8 |
| 12.2 | Ast | Nom | 20:16:57 | 16183 | 21:47:29 | 352 | 91 | 2.89 | 2.24 |
| 13.2 | Ast | H | 19:59:27 | 16260 | 20:25:05 | 244 | 25 | 11.67 | 10.3 |
| 14.1 | G | Nom | 20:07:39 | 18861 | 22:09:33 | 542 | 121 | 2.29 | 1.96 |
| 14.2 | Ast | H | 20:10:58 | 17180 | 20:44:18 | 621 | 33 | 7.60 | 15.4 |
| 14.3 | Aug | Nom | 20:13:43 | 20014 | 22:26:17 | 621 | 132 | 2.44 | 1.79 |
| 15.1 | Aug | Nom | 19:54:30 | 17388 | 21:35:08 | 623 | 101 | 2.75 | 2.07 |
| 15.3 | Aug | Nom | 20:07:17 | 12831 | 21:23:23 | 583 | 76 | 2.60 | 2.14 |
| 16.1 | Aug | H | 19:31:46 | 17289 | 20:13:05 | 324 | 94 | 8.08 | 6.52 |
| 16.2 | Aug | Nom | 19:42:40 | 16143 | 21:16:52 | 326 | 40 | 2.60 | 2.18 |
| 17.1 | Aug | Nom | 19:27:37 | 16633 | 21:10:20 | 260 | 103 | 2.49 | 1.90 |
| 18.1 | Aug | Nom | 19:18:26 | 16539 | 21:00:01 | 430 | 102 | 2.52 | 1.76 |
| 18.2 | Aug | Nom | 19:19:28 | 16438 | 20:59:31 | 430 | 100 | 2.56 | 1.88 |